\documentclass[twocolumn,groupedaddress,showpacs,nofootinbib]{revtex4}
\usepackage{graphicx}
\usepackage{amsmath}
\usepackage{amssymb}
\usepackage{epsfig}

\begin{document}

\title{Realistic Neutrinogenesis with Radiative Vertex Correction}

\author{Pei-Hong Gu$^{1}_{}$}
\email{pgu@ictp.it}

\author{Hong-Jian He$^{2}_{}$}
\email{hjhe@tsinghua.edu.cn}

\author{Utpal Sarkar$^{3}_{}$}
\email{utpal@prl.res.in}

\affiliation{
$^{1}_{}$The Abdus Salam International Centre for
Theoretical Physics, Strada Costiera 11, 34014 Trieste, Italy
\\
$^{2}_{}$Center for High Energy Physics,
Tsinghua University, Beijing 100084, China
\\
$^{3}_{}$Physical Research Laboratory,
Ahmedabad 380009, India}

 \begin{abstract}
 We propose a new model for naturally realizing light Dirac neutrinos
 and explaining the baryon asymmetry of the universe through
 neutrinogenesis. To achieve these, we present a minimal construction
 which extends the standard model with a real singlet scalar,
 a heavy singlet Dirac fermion and a heavy doublet scalar besides
 three right-handed neutrinos, respecting lepton number conservation
 and a $Z_2^{}$ symmetry. The neutrinos acquire small Dirac masses
 due to the suppression of weak scale over a heavy mass scale.
 As a key feature of our construction, once the heavy Dirac fermion and
 doublet scalar go out of equilibrium, their decays induce the CP
 asymmetry from the interference of tree-level processes with the
 \textit{radiative vertex corrections} (rather than the self-energy
 corrections). Although there is no lepton number violation, an equal and
 opposite amount of CP asymmetry is generated in the left-handed and the
 right-handed neutrinos. The left-handed lepton asymmetry would then be
 converted to the baryon asymmetry in the presence of the sphalerons,
 while the right-handed lepton asymmetry remains unaffected.
 \\ [2mm]
 \pacs{98.80.-k, 11.30.Er, 14.60.Pq \hfill
 arXiv:\,0709.1019 [hep-ph]}
 \end{abstract}

 \maketitle

 Strong evidences from neutrino oscillation
 experiments\,\cite{pdg2006} so far have pointed to tiny
 but nonzero masses for active neutrinos.
 The smallness of the neutrino masses can be elegantly
 understood via seesaw mechanism\,\cite{minkowski1977}
 in various extensions of the standard model (SM).
 The origin of the observed baryon asymmetry\,\cite{pdg2006}
 in the universe poses a real challenge to the SM, but within
 the seesaw scenario, it can be naturally explained through
 leptogenesis \cite{fy1986,luty1992,fps1995,ms1998,di2002,kl1984}.

 In the conventional leptogenesis scenario, the lepton number
 violation is essential as it is always associated with the
 mass-generation of Majorana neutrinos.
 However, the Majorana or Dirac nature of the neutrinos
 is unknown a priori and is awaiting for the upcoming
 experimental determination.
 It is important to note\,\cite{ars1998,dlrw1999} that
 even with lepton number conservation, it is
 possible to generate the observed baryon asymmetry in the universe.
 Since the sphaleron processes\,\cite{krs1985} have no
 direct effect on the right-handed fields, a nonzero lepton
 asymmetry stored in the left-handed fields, which is equal
 but opposite to that stored in the right-handed fields,
 can be partially converted to
 the baryon asymmetry as long as the interactions between
 the left-handed lepton number and the right-handed lepton
 number are too weak to realize an equilibrium before the
 electroweak phase transition, the sphalerons convert the
 lepton asymmetry in the left-handed fields, leaving the
 asymmetry in the right-handed fields unaffected \cite{dlrw1999,mp2002,ap2006,gdu2006,gh2006}.

 For all the SM species, the Yukawa interactions are
 sufficiently strong to rapidly cancel the stored left- and
 right-handed lepton asymmetry. However, the effective Yukawa
 interactions of the ultralight Dirac neutrinos are
 exceedingly weak\,\cite{rw1983,rs1984}
 and thus will not reach equilibrium until the temperatures
 fall well below the weak scale.
 In some realistic models\,\cite{mp2002,gdu2006,gh2006},
 the effective Yukawa couplings of the Dirac neutrinos are
 naturally suppressed by the ratio of the weak scale over the
 heavy mass scale. Simultaneously, the heavy particles can
 decay with the CP asymmetry to generate the expected
 left-handed lepton asymmetry after they are
 out of equilibrium. This new type of leptogenesis
 mechanism is called neutrinogenesis \cite{dlrw1999}.

 In this paper, we propose a new model to generate the
 small Dirac neutrino masses and explain the origin of
 cosmological baryon asymmetry, by extending the SM with
 a real scalar, a heavy Dirac fermion singlet and a heavy doublet
 scalar besides three right-handed neutrinos.
 In comparison with all previous
 realistic neutrinogenesis
 models \cite{mp2002,gdu2006,gh2006}, the
 Dirac neutrino masses in our new model
 are also suppressed by the ratio of the weak
 scale over the heavy mass scale,
 but the crucial difference is that in the decays of the
 heavy particles, the \textit{radiative vertex corrections}
 (instead of the self-energy corrections)
 interfere with the tree-level diagrams
 to generate the required CP asymmetry
 and naturally realize neutrinogenesis.

\begin{table}
\begin{center}
\begin{tabular}{c|ccc}
\hline\hline
\\[-2mm]
~~Fields          \quad\quad & \quad~~\,$SU(2)_{L}^{}$ \quad\quad &
$ U(1)_{Y}^{} \quad\quad $ & $  Z_{2}^{} $~~\quad\quad
\\[1.5mm]
\hline
\\[-2mm]
~~$\psi_{L}^{}$  \quad\quad &     ~~\,\textbf{2}    \quad\quad & $
-1/2 \quad\quad $ & $ + $\quad\quad
\\[1.5mm]
~~$\phi$         \quad\quad &     ~~\,\textbf{2}    \quad\quad & $
-1/2 \quad\quad $ & $  + $\quad\quad
\\[1.5mm]
~~$\nu_{R}^{}$   \quad\quad &     ~~\,\textbf{1}    \quad\quad & $
~\,0 \quad\quad $ &  $  - $\quad\quad
\\[1.5mm]
~~$D_{L,R}^{}$         \quad\quad & ~~\,\textbf{1} \quad\quad & $
~\,0 \quad\quad $ &  $ + $\quad\quad
\\[1.5mm]
~~$\eta$         \quad\quad &     ~~\,\textbf{2} \quad\quad & $
-1/2 \quad\quad $ &  $ - $\quad\quad
\\[1.5mm]
~~$\chi$         \quad\quad &     ~~\,\textbf{1}    \quad\quad & $
~\,0 \quad\quad $ & $  -$\quad\quad
\\[-2mm]
\\ \hline \hline
\end{tabular}
\caption{The field content of our model, where
 $\psi_{L}^{}$ is the left-handed lepton doublet,
 $\phi$ is the SM Higgs doublet, $\nu_{R}^{}$ is the
 right-handed neutrino, $D_{L,R}^{}$ is the heavy singlet Dirac
 fermion, $\eta$ is the heavy doublet scalar,
 and $\chi$ is the real scalar.
 For simplicity the family indices and
 the other SM fields (carrying even $Z_2^{}$ parity)
 are omitted from the Table.}
 \label{charge}
 \end{center}
 \end{table}

\begin{figure*}
\vspace{5.5cm} \epsfig{file=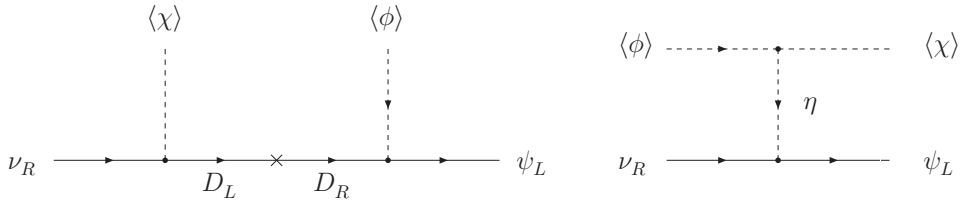, bbllx=6cm, bblly=6.0cm,
bburx=16cm, bbury=16cm, width=7cm, height=7cm, angle=0, clip=0}
\vspace{-9cm} \caption{\label{massgeneration} The neutrino
mass-generation. The left diagram is the type-I Dirac seesaw while
the right one is the type-II Dirac seesaw.}
\end{figure*}

 We summarize the field content in Table\,\ref{charge},
 in which $\psi_{L}^{}$, $\phi$, $\nu_{R}^{}$, $D_{L,R}^{}$,
 $\eta$ and $\chi$ denote the left-handed lepton doublets,
 the SM Higgs doublet, the right-handed neutrinos, the heavy
 singlet Dirac fermion, the heavy doublet scalar and the
 real scalar, respectively. Here
 $\psi_{L}^{}$, $\nu_{R}^{}$, $D_L^{}$ and $D_R^{}$ carry lepton
 number $1$ while $\phi$, $\eta$ and $\chi$ have zero lepton number.
 For simplicity, we have omitted the family indices as well
 as other SM fields, which carry even
 parity under the discrete symmetry $Z_{2}^{}$.
 It should be noted that the conventional dimension-4
 Yukawa interactions among the left-handed lepton doublets,
 the SM Higgs doublet and the right-handed neutrinos are
 forbidden under the $Z_{2}^{}$ symmetry.
 Our model also exactly conserves the lepton number,
 so we can write down the relevant Lagrangian
 as below,
\begin{eqnarray}
\label{lagrangian1}
 -\mathcal{L}
 &\supset&
 \left\{f_{i}^{}\overline{\psi_{Li}^{}}\phi D_{R}^{}
 + g_{i}^{}\chi
 \overline{D_{L}^{}}\nu_{Ri}^{}
 +y_{ij}^{}\overline{\psi_{Li}^{}}\eta\nu_{Ri}^{}
 -\mu\chi\eta^{\dagger}_{}\phi\right.\nonumber\\
 &&
 \left.+M_{D}^{}\overline{D_{L}^{}}D_{R}^{}
 +\textrm{h.c}\right\}+M_{\eta}^{2}\eta^{\dagger}_{}\eta\,,
 \end{eqnarray}
 where $f_i^{}$,\, $g_i^{}$ and $y_{ij}^{}$
 are the Yukawa couplings, while the
 cubic scalar coupling $\mu$ has mass-dimension equal one.
 The parameters $M_D$ and $M_\eta$ in (\ref{lagrangian1})
 are the masses of the heavy singlet fermion $D$ and the
 heavy Higgs doublet $\eta$, respectively.
 Note that in the Higgs potential the scalar doublet
 $\eta$ has a positive mass-term as shown in the above
 Eq.\,(\ref{lagrangian1}), while the Higgs doublet $\phi$ and
 singlet $\chi$ both have negative mass-terms\,\footnote{The
 general Higgs potential $V(\phi,\eta,\chi)$ was given in the
 Appendix of our first paper in Ref.\,\cite{gh2006}.}.

 The lepton number conservation ensures that there is no
 Majorana mass term for all fermions. As we will discuss
 below, the vacuum expectation value (\textit{vev}) of
 $\eta$ comes out to be much less than the \textit{vev}  of
 the other fields.
 Thus the first two terms generate mixings
 of the light Dirac neutrinos with the heavy Dirac fermion,
 while the third term gives the light Dirac neutrino mass
 term. The complete mass matrix can now be written in the
 basis
 $\left\{ \nu_L^{},~ D_L^{},~\nu_R^{},~D_R^{}\right\}$ as
 \begin{eqnarray}
 \label{eq:Mnu44}
 M = \left[ \begin{array}{cccc}0 & 0 & a & b
 \\ 0 & 0 & c & d
 \\ a^{\dagger}_{} & c^{\dagger}_{} & 0 & 0
 \\ b^{\dagger}_{} & d^{\dagger}_{} & 0 &
 0\end{array}\right] \,,
 \end{eqnarray}
 where \,$a \equiv y\langle \eta \rangle$,\,
 $ b\equiv f \langle \phi\rangle$,\,
 $ c\equiv g \langle \chi \rangle$\, and
 \,$d\equiv M_{D}^{}$.\,
 As will be shown below, $\,d\gg a,b,c\,$.\,
 So, the diagonalization of the mass matrix
 (\ref{eq:Mnu44}) generates
 the light Dirac neutrino masses of order
 \,$a - bc/d$\, and a heavy Dirac fermion
 mass of order \,$d$\,.

 As shown in Fig.\,\ref{massgeneration},
 at low energy we can
 integrate out the heavy singlet fermion as well as the
 heavy doublet scalar.
 Then we obtain the following
 effective dimension-5 operators,
 \begin{eqnarray}
 \label{operator}
 \mathcal{O}_{5}^{}
 =\frac{\,f_{i}^{}g_{j}^{}\,}{M_{D}^{}}
 \overline{\psi_{Li}^{}}\phi\nu_{Rj}^{}\chi
 -\frac{\,\mu y_{ij}^{}\,}{M_{\eta}^{2}}
 \overline{\psi_{Li}^{}}\phi\nu_{Rj}^{}\chi
 +\textrm{h.c.}\,.
 \end{eqnarray}
 Therefore, once the SM Higgs doublet $\phi$ and
 the real scalar $\chi$ both acquire their \textit{vev}s,
 the neutrinos naturally acquire small Dirac masses,
 \begin{eqnarray}
 \mathcal{L}_m &=&
 -\left(m_{\nu}^{}\right)_{ij}^{}
 \overline{\nu_{Li}^{}}\nu_{Rj}^{}+\textrm{h.c.}\,,
 \end{eqnarray}
 where
 \begin{eqnarray}
 \label{diracmass}
 m_{\nu}^{}
 ~\equiv~ m_{\nu}^{I}+ m_{\nu}^{II} \,,
 \end{eqnarray}
 with \cite{rw1983}
 \begin{eqnarray}
 \label{diracmass1}
 &&
 \left(m_{\nu}^{I}\right)_{ij}^{} ~=~
 -f_{i}^{}g_{j}^{}\frac{\langle \phi \rangle \langle \chi
 \rangle}{M_{D}^{}}
 ~=~ -\frac{\,(bc)_{ij}^{}\,}{d}
 \,,
 \end{eqnarray}
 and \cite{gh2006}
 \begin{eqnarray}
 \label{diracmass2}
 &&
 \left(m_{\nu}^{II}\right)_{ij}^{~}
 ~=~
 y_{ij}^{}\frac{\mu\langle \phi \rangle \langle \chi
 \rangle}{M_{\eta}^{2}}
 ~=~ a_{ij}^{~}
 \,.
 \end{eqnarray}
 To quantify the second equality in (\ref{diracmass2}),
 we note that different from the SM Higgs doublet,
 the heavy scalar doublet $\eta$
 has a positive mass-term in the Higgs potential, so it
 will develop a tiny nonzero \textit{vev}
 until $\phi$ and $\chi$ both acquire
 their \textit{vev}s \cite{gh2006},
 \begin{eqnarray}
 \label{doubletvev}
 \langle\eta\rangle
 & \simeq &
 \frac{\,\mu\langle \phi\rangle \langle \chi \rangle \,}
 {M^{2}_{\eta}}\,.
 \end{eqnarray}
 With this we can derive the neutrino mass formula
 ~$m_{\nu}^{II}= y\langle \eta \rangle \equiv a$~ from
 the Lagrangian (\ref{lagrangian1}), which
 confirms the Eq.\,(\ref{diracmass2}) above.
 In the reasonable parameter space of
 $\,M_D^{} \sim M_{\eta}^{}\sim \mu
   \gg \left<\chi\right>,\left<\phi\right> $\,
 and $\,(f,\,g,\,y)={O}(1)$\,,\,
 we can naturally realize
 \,$ a \ll b,c \ll d\,$.\,
 Furthermore, using the second relations in
 (\ref{diracmass1}) and (\ref{diracmass2}) we can
 re-express the summed neutrino mass matrix as
  \begin{eqnarray}
 \label{diracmass}
 m_{\nu}^{}
 ~\equiv~ m_{\nu}^{I}+ m_{\nu}^{II} = -bc/d + a \,.
 \end{eqnarray}
 This is consistent with the direct diagonalization
 of the original Dirac mass matrix (\ref{eq:Mnu44}),
 which we have mentioned below (\ref{eq:Mnu44}).

 It is clear that this mechanism of the neutrino mass
 generation has two essential features:
 (i) it generates Dirac masses for neutrinos, and
 (ii) it retains the essence of the conventional seesaw
 \cite{minkowski1977} by making the neutrino masses tiny
 via the small ratio of the weak scale over the heavy mass
 scale. It is thus called Dirac Seesaw \cite{gh2006}.
 In particular, compared to the classification of the
 conventional type-I and type-II seesaw, we may refer to
 Eqs.\,(\ref{diracmass1}) and (\ref{diracmass2}) as the
 type-I and type-II Dirac seesaw, respectively.

 From Eq.\,(\ref{diracmass}) we see
 that both type-I and type-II seesaws
 can contribute to the $3\times 3$ mass-matrix $m_\nu^{}$
 for the light neutrinos.
 There are three possibilities in general:
 (i) $m^I_\nu \gg m^{II}_\nu$, or
 (ii) $m^{I}_\nu \sim m^{II}_\nu$, or
 (iii) $m^{I}_\nu \ll m^{II}_\nu$.
 We note that for case-(iii), the type-II
 contribution alone can accommodate the neutrino oscillation data
 even if type-I is fully negligible; while for case-(i) and -(ii),
 the type-II contribution
 should still play a nontrivial role for $\nu$-mass generation
 because $m^I_\nu$ is rank-1 and additional contribution from
 $m^{II}_\nu$ is necessary.
 The rank-1 nature of \,$m_{\nu}^{I}=bc/d$\, is due to that
 there is only one singlet heavy fermion in our current minimal
 construction, which means that \,$m_{\nu}^{I}$\, has two vanishing
 mass-eigenvalues. Hence, to accommodate
 the neutrino oscillation data\,\cite{pdg2006}
 in the case-(i) and -(ii) of
 our minimal construction always requires
 nonzero contribution \,$m_{\nu}^{II}$\, from
 the type-II Dirac seesaw\,\footnote{
 Note that the type-I Dirac seesaw alone can accommodate the
 oscillation data once we extend the current minimal construction
 to include a second heavy fermion $D'$
 which makes $m_{\nu}^{I}$ rank 2; this is similar to the
 minimal (Majorana) neutrino seesaw studied
 before\,\cite{MMnuSS}.}.\,
 Let us explicitly analyze how this can be realized
 for the case-(i) and -(ii).
 As $m_{\nu}^{I}$ is rank-1,
 we can consider a basis for \,$m_{\nu}^{I}$\,
 where one of the two massless states is manifest, i.e.,
 \,$b_1^{} = c_1^{} =0$\,.\,
 For the remaining components of $b$ and $c$,
 we choose a generic parameter set\footnote{
 Here we consider the difference between any two
 of the four components $|b_j^{}|$ and $|c_j^{}|$
 ($j=2,3$) to be much smaller themselves.},
 $\,b_2^{} \approx b_3^{} \approx -c_2^{} \approx -c_3^{}\,$,\,
 which naturally realizes the
 maximal mixing angle \,$\theta_{23}^{}=45^{\circ}_{}$\,
 for explaining the atmospheric neutrino mixing.
 Including the type-II Dirac-seesaw matrix
 \,$m_{\nu}^{II}=a$\,
 will then account for the other mixing angles
 ($\theta_{12}^{},\,\theta_{13}^{}$) and
 the two other neutrino masses. Thus, we can naturally
 realize the light neutrino mass-spectrum via both
 normal hierarchy (NH) and inverted hierarchy (IH) schemes.
 To be concrete, the NH-scheme is realized in our case-(i) where
 the type-II Dirac-seesaw matrix
 \,$m_{\nu}^{II}=a\equiv \delta \ll m_0^{}
    \sim m_{\nu}^{I}$\,,\,
 with $m_0^{}$ the neutrino mass scale
 (fixed by the atmospheric neutrino mass-squared-difference
 $\Delta_{\textrm{a}}^{}$ with
 $\,m_0^{} \equiv \sqrt{\Delta_{\textrm{a}}^{}}\,$) and its relations
 to the nonzero $(b_j^{},\,c_j^{}$) are defined via
 $\,b_j^{} = \sqrt{m_0^{}d/2}+{O}(\delta)$ and
 $\,c_j^{} = -\sqrt{m_0^{}d/2}+{O}(\delta)$ for $j=2,3$.\,
 Thus we have
 \begin{eqnarray}
 \label{NH}
 \displaystyle
 m_{\nu}^{} = -\frac{bc}{d} + a
 =m_0^{}\left(\begin{array}{ccc}
 0 & 0 & 0\\
 0 & \frac{1}{2} & \frac{1}{2}\\
 0 & \frac{1}{2} & \frac{1}{2}
 \end{array}\right) + {O}(\delta )\,.
 \end{eqnarray}
 It is clear that Eq.\,(\ref{NH}) predicts
 the neutrino masses,
 $\,(m_1^{},\,m_2^{},\,m_3^{})
  = m_0^{}(0,\,0,\,1) + {O}(\delta)\,$,
 consistent with the NH mass-spectrum.
 Next, the IH-scheme can be realized in our case-(ii) where
 the type-II Dirac-seesaw matrix
 \,$m_{\nu}^{II}=a\equiv
 m_0^{}{\rm diag}(1,0,0) + \delta \,\sim\, m_\nu^I $\, with
 $\,\delta \ll m_0^{}\,$,\, while the structure of
 the type-I Dirac seesaw matrix \,$m_{\nu}^{I}$\,
 remains the same,
 \begin{eqnarray}
 \label{IH}
 \displaystyle
 m_{\nu}^{} = -\frac{bc}{d} + a
 =m_0^{}\left(\begin{array}{ccc}
 1 & 0 & 0\\
 0 & \frac{1}{2} & \frac{1}{2}\\
 0 & \frac{1}{2} & \frac{1}{2}
 \end{array}\right) + {O}(\delta )\,.
 \end{eqnarray}
 From this equation we deduce the neutrino masses,
 $\,(m_1^{},\,m_2^{},\,m_3^{})
  = m_0^{}(1,\,1,\,0) + {O}(\delta)\,$,
 consistent with the IH mass-spectrum.
 So far we have discussed all three possibilities, the case-(i),
 -(ii) and -(iii), regarding the relative contributions of the
 type-I versus type-II seesaw to the neutrino mass matrix
 $m_\nu^{}$ for accommodating the oscillation data.
 The question of which one among these three possibilities is
 realized in nature should be answered by a more fundamental theory
 which can precisely predict the Yuakawa couplings and masses
 for $D$ and $\eta$ as well as the {\it vev} of \,$\chi$\,.\,
 Finally, we also note that as the neutrinos being Dirac particles,
 our Dirac-seesaw construction will be consistent with
 the possible non-observation of the neutrinoless double
 beta decays ($0\nu\beta\beta$) which will be tested in
 the upcoming $0\nu\beta\beta$-experiments\,\cite{0nu2beta}.

 The real scalar $\chi$ is expected to acquire its \textit{vev} near
 the weak scale, so we will set \,$ \langle\chi\rangle$\, around
 ${O}(\textrm{TeV})$ \footnote{Here we comment on the cosmological
 domain wall problem associated with spontaneous breaking
 of a discrete $Z_2$ symmetry. This problem arises during the
 phase transition (when the broken discrete symmetry gets
 restored at the transition temperature) because of the production of
 topological defects -- domain walls
 which carry too much energy and trouble
 the standard big-bang cosmology\,\cite{DW}.
 This can be avoided by inflation as long as phase transition
 temperature is above the inflation scale\,\cite{kt1990,Linde}.
 Another resolution\,\cite{dvali1} to the domain wall problem is
 realized by the possibility of symmetry non-restoration at high
 temperature\,\cite{Weinberg}. It is also very possible
 that a discrete symmetry like $Z_2$ is not a basic symmetry but
 appears as a remnant of a continuous symmetry such as $U(1)$ which
 is free from the domain wall problem\,\cite{dvali2}.
 Finally, the $Z_{2}^{}$ symmetry in our model can also be replaced by
 a global $U(1)_{D}^{}$ as in the pure
 type-I Dirac seesaw model\,\cite{rw1983}, and
 the phenomenology of the Goldstone boson associated
 with this $U(1)_{D}^{}$ breaking
 was discussed in \cite{gdu2006}.}.
 Under this setup, it is straightforward
 to see that $m_{\nu}^{I}$ will be efficiently suppressed by the
 ratio of the weak scale over the heavy mass. For instance, we find
 that \,$m_{\nu}^{I}={O}(0.1)\,\textrm{eV}$\, for \,$\,M_{D}^{}=
 {O}(10^{13-15}_{})\,\textrm{GeV}$\, and \,$(f,\, g,\,y) =
 {O}(0.1-1)$\,,\, where \,$\langle \phi\rangle\simeq
 174\,\textrm{GeV}$.\, It also is reasonable to set the trilinear
 scalar coupling $|\mu|$ to be around the scale of the $\eta$ mass
 $M_{\eta}^{}$.\, In consequence, the neutrino mass $m_{\nu}^{II}$
 in (\ref{diracmass2}) will be highly suppressed, similar to
 $m_{\nu}^{I}$. For example, we derive
 \,$m_{\nu}^{II}={O}(0.1)\,\textrm{eV}$\, for \,$M_{\eta}^{}
        ={O}(10^{13-15}_{})\,\textrm{GeV}$\, and
 \,$(y,\,\mu/M_\eta ) ={O}(0.1-1)$\,.\,
 So, we can naturally realize the Dirac neutrino masses
 around \,${O}(0.1)\,\textrm{eV}$\,.

 We now demonstrate how to generate the observed baryon
 asymmetry in our model by invoking the neutrinogenesis
 \cite{dlrw1999} mechanism. Since the sphaleron processes
 \cite{krs1985} have no direct effect on the right-handed
 neutrinos, and the effective Yukawa interactions of the
 Dirac neutrinos are too weak to reach the equilibrium
 until temperatures fall well below the weak scale, the
 lepton asymmetry stored in the left-handed leptons,
 which is equal but opposite to that stored in the
 right-handed neutrinos, can be partially converted to the
 baryon asymmetry by sphalerons. In particular, the final
 baryon asymmetry should be
 \begin{eqnarray}
 B ~=~ \frac{28}{79}\left(B-L_{SM}^{}\right)
   ~=~ -\frac{28}{79}L_{SM}^{} \,,
 \end{eqnarray}
 for the SM with three generation fermions
 and one Higgs doublet.

 In the pure type-I Dirac seesaw scenario \cite{gdu2006},
 we can generate the CP asymmetry through the interferences
 between the tree-level decay and the self-energy loops if
 there exist at least two heavy fermion singlets. Similarly,
 the pure type-II Dirac seesaw model \cite{gh2006} also needs
 two heavy scalar doublets to obtain the self-energy
 loops in the decays. In the following, we
 shall focus on the minimal construction
 with only one heavy singlet
 fermion and one heavy doublet scalar to realize
 the radiative vertex corrections for
 the CP asymmetry, although further extensions are allowed
 in our current scenario.

\begin{figure*}
\vspace{7cm} \epsfig{file=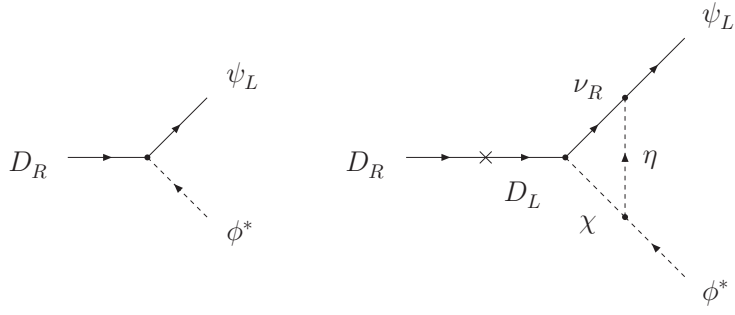, bbllx=5cm, bblly=6.0cm,
bburx=15cm, bbury=16cm, width=7.5cm, height=7.5cm, angle=0, clip=0}
\vspace{-9.5cm}
 \caption{\label{decay1} The heavy singlet Dirac fermion
 decays to the left-handed leptons and the SM Higgs boson
 at one-loop order. }
\end{figure*}

 \begin{figure*}
 \vspace{8cm} \epsfig{file=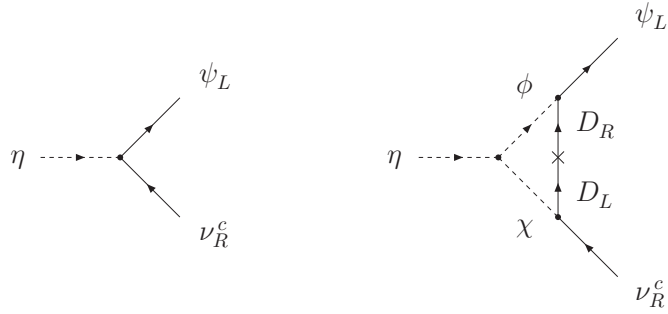, bbllx=5cm, bblly=6.0cm,
 bburx=15cm, bbury=16cm, width=7.5cm, height=7.5cm, angle=0, clip=0}
 \vspace{-9.5cm}
 \caption{\label{decay2} The heavy doublet scalar
 decays to the leptons at one-loop order. }
 \end{figure*}

 In this framework, depending on the values of the
 masses and couplings, the leptogenesis can be realized
 either from the decay of the heavy singlet fermion or from
 the decay of the heavy doublet scalar. From the decay of
 the heavy singlet fermion to the left-handed leptons
 and the SM Higgs doublet, as shown in
 Fig.\,\ref{decay1}, the CP asymmetry is given by
 \begin{eqnarray}
 \label{cp1} \varepsilon^{I}_{}
 &\equiv&
 \frac{\Gamma\left(D_R^{}\rightarrow\psi_{L}^{}\phi^{\ast}_{}
 \right)
 -\Gamma\left(D^c_R\rightarrow\psi_{L}^{c}\phi\right)}
 {\Gamma_{D}^{}}\nonumber
 \\[3mm]
 &=&
 \frac{1}{4\pi}\frac{\textrm{Im}
 \left[\textrm{Tr}\left(f^{\dagger}_{}yg^{\dagger}_{}\right)
 \mu\right]M_{\eta}^{2}}
 {\left[\textrm{Tr}
 \left(f_{}^{\dagger}f\right)+\frac{1}{2}\textrm{Tr}
 \left(g_{}^{\dagger}g\right)\right]M_{D}^{3}}\nonumber
 \\[1.7mm]
 &&
 \times
 \ln\left(1+\frac{M_{D}^{2}}{M_{\eta}^{2}}
 \right)\,,
 \end{eqnarray}
 where
 \begin{eqnarray}
 \Gamma_{D}^{} &=&
 \frac{1}{16\pi}\left[\textrm{Tr}
 \left(f_{}^{\dagger}f\right)+\frac{1}{2}\textrm{Tr}
 \left(g_{}^{\dagger}g\right)\right]M_{D}^{}
 \end{eqnarray}
 is the total decay width of \,$D$\, or \,$D_{}^c$\,.
 Here we have taken
 $M_{D}^{}$ to be real after proper phase rotation.
 Furthermore, from the decay of the heavy doublet scalar
 to the left-handed leptons and the right-handed
 neutrinos, a CP asymmetry can also be produced.
 It is given by the interference of the tree-level process
 with the one-loop vertex diagram as shown in
 Fig.\,\ref{decay2},
\begin{eqnarray}
 \label{cp2}
 \hspace*{-3mm}
 \varepsilon^{II}_{}
 &\equiv&
 \frac{\,\Gamma\left(\eta\rightarrow\psi_{L}^{}\nu_{R}^{c}
 \right)
 -\Gamma\left(\eta^{\ast}_{}\rightarrow\psi_{L}^{c}
 \nu_{R}\right)\,}
 {\Gamma_{\eta}^{}}
 \nonumber\\[2mm]
 \hspace*{-3mm}
 &=&
 \frac{1}{4\pi}\frac{\textrm{Im}\left[\textrm{Tr}
 \left(f^{\dagger}_{}yg^{\dagger}_{}\right)
 \mu\right]M_{D}^{}}{\left[\textrm{Tr}
 \left(y_{}^{\dagger}y\right)M_{\eta}^{2}+
 |\mu|^{2}_{}\right]}
 \ln\!\left(\!1+\frac{M_{\eta}^{2}}{M_{D}^{2}}\right)
 \end{eqnarray}
 where
 \begin{eqnarray}
 \Gamma_{\eta}^{} ~=~
 \frac{1}{16\pi}
 \left[\textrm{Tr}\left(y^{\dagger}_{}y\right)
 +\frac{\left|\mu\right|^{2}_{}}{M_{\eta}^{2}}\right]
 M_{\eta}^{}\,
 \end{eqnarray}
 is the total decay width of
 \,$\eta$\, or \,$\eta^{\ast}_{}$\,.

 In the case where the masses of
 the heavy singlet fermion
 and heavy doublet scalar locate around the same scale,
 and also their couplings are of
 the same order of magnitude,
 the two types of asymmetry
 of Eqs.\,(\ref{cp1}) and (\ref{cp2}) can be both
 important for the neutrinogenesis.
 For illustration below,
 we will analyze two typical scenarios
 where one process dominates over the other.

 Scheme-1 is defined for
 \,$M_{D}^{}\ll M_{\eta}^{}$\, and
 \,$f\sim g\sim y$,\,
 under which the final left- or right-handed lepton
 asymmetry mainly comes from the pair decays of
 $(D,\,D^{c}_{})$. We can simplify the CP
 asymmetry (\ref{cp1}) as
 \begin{eqnarray}
 \hspace*{-5mm}
 \varepsilon^{I}_{}
 &\simeq&\frac{1}{64\pi^{2}_{}}
 \frac{M_{D}^{}M_{\eta}^{2}\textrm{Im}
 \left[\textrm{Tr}\left(m_{\nu}^{I\dagger}m_{\nu}^{II}
 \right)\right]}
 {\langle\phi\rangle^{2}_{}\langle\chi\rangle^{2}_{}
 \Gamma_{D}^{}}
 \nonumber
 \\[3mm]
 \hspace*{-5mm}
 &=&
 \left[\frac{45}{\left(4\pi\right)^{7}_{}
 g_{\ast}^{}}\right]^{\frac{1}{2}}_{}
 \frac{1}{K_{D}^{}}\frac{M_{\eta}^{2}}{M_{D}^{2}}
 \nonumber
 \\[1.8mm]
 &&
 \times\frac{M_{\textrm{Pl}}^{}M_{D}^{}\textrm{Im}
 \left[\textrm{Tr}\left(m_{\nu}^{I\dagger}m_{\nu}^{II}
 \right)\right]}
 {\langle\phi\rangle^{2}_{}\langle\chi\rangle^{2}_{}}
 \end{eqnarray}
 with
 \begin{eqnarray}
 \label{kd} K_{D}^{} &\equiv&
 \left.\frac{\Gamma_{D}^{}}{H}\right|^{}_{T=M_{D}^{}}
 \end{eqnarray}
 as a measurement of the deviation from equilibrium
 for $D$. Here $H$ is the Hubble constant,
 \begin{eqnarray}
 \label{eq:HubbleC}
 H(T) &=&
 \left(\frac{4\pi^{3}_{}g_{\ast}^{}}{45}
 \right)^{\frac{1}{2}}
 \frac{T^{2}_{}}{M_{\textrm{Pl}}^{}}\,,
 \end{eqnarray}
 with \,$g_{\ast}^{}={O}(100)$\, and
 \,$M_{\textrm{Pl}}\simeq 1.2\times
    10^{19}_{}\,\textrm{GeV}$.\,
 Note that there is a correlation between $K_{D}^{}$ and
 $m_{\nu}^{I}$,
 \begin{eqnarray}
 \overline{m}_{I}^{2}&\equiv&
 \textrm{Tr}\left(m_{\nu}^{I\dagger}m_{\nu}^{I}\right)
 \nonumber\\
 &=&
 \textrm{Tr}\left(f^{\dagger}_{}fgg^{\dagger}_{}\right)
 \frac{\langle\phi\rangle^{2}_{}
 \langle\chi\rangle^{2}_{}}{M_{D}^{2}}
 \nonumber\\
 &=&\sum_{i}^{}\left|f_{i}^{}
 \right|_{}^{2}\left|g_{i}^{}\right|_{}^{2}
 \frac{\langle\phi\rangle^{2}_{}
 \langle\chi\rangle^{2}_{}}{M_{D}^{2}}
 \nonumber\\
 &<&\sum_{i}^{}\left|f_{i}^{}
 \right|_{}^{2}\sum_{j}^{}\left|g_{j}^{}\right|_{}^{2}
 \frac{\langle\phi\rangle^{2}_{}
 \langle\chi\rangle^{2}_{}}{M_{D}^{2}}
 \nonumber\\
 &=&
 2\left(16\pi\right)^{2}_{}
 B_{L}^{}B_{R}^{}\Gamma_{D}^{2}
 \frac{\langle\phi\rangle^{2}_{}
 \langle\chi\rangle^{2}_{}}{M_{D}^{4}}
 \nonumber\\
 &=&
 \frac{2\left(4\pi\right)^{5}_{}g_{\ast}^{}}{45}
 B_{L}^{}B_{R}^{}K_{D}^{2}
 \frac{\langle\phi\rangle^{2}_{}
 \langle\chi\rangle^{2}_{}}{M_{\textrm{Pl}}^{2}}\,,
 \end{eqnarray}
 and hence
 \begin{eqnarray}
 \label{constrain}
 K_{D}^{} & > &
 \left[\frac{45}{2\left(4\pi\right)^{5}_{}
 g_{\ast}^{}B_{L}^{}B_{R}^{}}\right]^{\frac{1}{2}}_{}
 \frac{\,M_{\textrm{Pl}}^{}\overline{m}_{I}\,}
 {\langle\phi\rangle\langle\chi\rangle}\,.
 \end{eqnarray}
 Here $B_{L}^{}$ and $B_{R}^{}$ are
 the branching ratios of the heavy fermion singlet
 decaying into the left-handed lepton doublets and
 the right-handed neutrinos, respectively.
 They satisfy the following relationship,
 \begin{eqnarray}
 \label{relation}
 B_{L}^{}+B_{R}^{}\equiv 1\,,\,  ~&\Rightarrow&~ \,
 B_{L}^{}B_{R}^{} \leqslant \frac{1}{4}\,.
 \end{eqnarray}

 For instance, we may choose the sample inputs,
 \,$\,M_{\eta}= 10\,M_{D}^{} = 1.8\times
 10^{12}_{}\,\textrm{GeV}\gg M_D^{}$,\,
 $\left<\phi\right>=174\,\textrm{GeV}$,\,
 $\left<\chi\right>=400\,\textrm{GeV}$\, and
 \,$(y,\,f,\, g,\,\mu/M_{\eta} ) \simeq
    (0.02,\,0.033,\,0.005,\,0.01)=O(0.01)$.\,
 Thus, we can estimate the light
 neutrino mass scale,
 \,$\overline{m}_{I}^{}
 ={O}(m_{\nu}^{I})=
 {O}(10\,m_{\nu}^{II})\simeq 0.06\,\textrm{eV}$.\,
 In consequence, we can estimate
 $\,B_{L}^{}B_{R}^{}\simeq 0.99\times 0.011
  \simeq 0.011$\,,\,
 and thus \,$K_{D}^{}\simeq 88$.\, This leads to
 \,$\varepsilon_{}^{I} \simeq -2.4\times 10^{-5}_{}$\,
 for the maximal CP phase. We then use the
 approximate relation \cite{kt1990,ht2001} to deduce
 the final baryon asymmetry,
 \begin{eqnarray}
 \label{asymmetry} Y_{B}^{} \equiv
 \frac{n_{B}^{}}{s}&\simeq&
 - \frac{28}{79} \times
 \frac{0.3\left(\varepsilon_{~}^{I}/g_{\ast}^{}\right)}
 {K_{D}^{}\left(\ln K_{D}^{}\right)^{0.6}_{}}
 \nonumber\\
 &\simeq&
 10^{-10}_{}\,.
 \end{eqnarray}
 This is consistent with
 the current observations \cite{pdg2006}.
 Furthermore, the relationship
 $m_{\nu}^{I}={O}(0.1\,\textrm{eV})
  \gg m_{\nu}^{II}$,\,
 shows the dominance of type-I Dirac seesaw.
 As we mentioned earlier, this
 can be realized via the NH mass-spectrum
 of the light neutrinos.

 We also note that even if
 the Dirac fermion singlet is at a
 fairly low mass scale, such as TeV,
 it is still feasible to
 efficiently enhance the CP asymmetry
 as long as the ratio
 \,${M_{\eta}}/{M_{D}}$\, is large enough.
 In other words, we can realize the low-scale
 neutrinogenesis without invoking the
 conventional resonant effect to enhance the CP
 asymmetry (which requires at
 least two heavy Dirac fermion singlets).

 Scheme-2 is defined for the other possibility with
 $M_{\eta}^{} \ll M_{D}^{}$ and
 $f\sim g\sim y$.
 Hence the final left- or right-handed lepton asymmetry
 is dominated by the pair decays of
 $(\eta,\,\eta^{\ast}_{})$.
 We derive the following CP asymmetry from (\ref{cp2}),
 \begin{eqnarray}
 \hspace*{-5mm}
 \varepsilon^{II}_{}
 &\simeq&
 \frac{1}{64\pi^{2}_{}}\frac{M_{\eta}^{3}\textrm{Im}
 \left[\textrm{Tr}\left(m_{\nu}^{I\dagger}m_{\nu}^{II}
 \right)\right]}
 {\langle\phi\rangle^{2}_{}\langle\chi\rangle^{2}_{}
 \Gamma_{\eta}^{}}
 \nonumber\\[3mm]
 \hspace*{-5mm}
 &=&\left[\frac{45}{\left(4\pi\right)^{7}_{}
 g_{\ast}^{}}\right]^{\!\frac{1}{2}}_{}\frac{1}{K_{\eta}^{}}\!\!
 \nonumber\\[1.8mm]
 && \times
 \frac{\,M_{\textrm{Pl}}^{}M_{\eta}^{}\textrm{Im}
 \left[\textrm{Tr}\left(m_{\nu}^{I\dagger}m_{\nu}^{II}
 \right)\right]\,}
 {\langle\phi\rangle^{2}_{}
  \langle\chi\rangle^{2}_{}}\,,
 \end{eqnarray}
 where $K_{\eta}^{}$ is given by
 \begin{eqnarray}
 K_{\eta}^{} &\equiv &
 \left.\frac{\Gamma_{\eta}^{}}{H}
 \right|^{}_{T=M_{\eta}^{}} \,,
 \end{eqnarray}
 with the Hubble constant $\,H(T)\,$ expressed in
 Eq.\,(\ref{eq:HubbleC}).
 Here the parameter $\,K_{\eta}^{}\,$
 measures the deviation from
 the equilibrium for \,$\eta$\,.
 We deduce the correlation between $K_{\eta}^{}$ and
 $m_{\nu}^{II}$,

 \begin{eqnarray}
 \overline{m}_{II}^{2}&\equiv&
 \textrm{Tr}\left(m_{\nu}^{II\dagger}m_{\nu}^{II}\right)
 \nonumber\\[3mm]
 &=& \textrm{Tr}\left(y^{\dagger}_{}y\right)
 \frac{|\mu|^{2}_{}\langle\phi\rangle^{2}_{}
 \langle\chi\rangle^{2}_{}}{M_{\eta}^{4}}
 \nonumber\\[3mm]
 &=&
 \left(16\pi\right)^{2}_{}B_{f}^{}B_{s}^{}
 \Gamma_{\eta}^{2}\frac{\langle\phi\rangle^{2}_{}
 \langle\chi\rangle^{2}_{}}{M_{\eta}^{4}}
 \nonumber
 \\[3mm]
 &=&
 \frac{\left(4\pi\right)^{5}_{}g_{\ast}^{}}{45}
 B_{f}^{}B_{s}^{}K_{\eta}^{2}
 \frac{\langle\phi\rangle^{2}_{}
 \langle\chi\rangle^{2}_{}}{M_{\textrm{Pl}}^{2}}\,,
 \end{eqnarray}
 and also
 \begin{eqnarray}
 K_{\eta}^{}=\left[\frac{45}{\left(4\pi\right)^{5}_{}
 g_{\ast}^{}B_{f}^{}B_{s}^{}}\right]^{\frac{1}{2}}_{}
 \frac{M_{\textrm{Pl}}^{}\overline{m}_{II}^{}}
 {\langle\phi\rangle\langle\chi\rangle}\,,
 \end{eqnarray}
 where $B_{f}^{}$ and $B_{s}^{}$ are the branching ratios
 of the heavy scalar doublet decaying into the light
 fermions and the scalars, respectively. Similar to
 Eq.\,(\ref{relation}), they satisfy
 \begin{eqnarray}
 B_{f}^{}+B_{s}^{}\equiv 1\,,\,
 ~&\Rightarrow&~ \, B_{f}^{}B_{s}^{} \leqslant
 \frac{1}{4}\,.
\end{eqnarray}

 For instance, given the sample inputs,
 $\,M_{\eta}^{}=26\mu =0.1\, M_{D}^{} =2\times
 10^{13}_{}\,\textrm{GeV}\ll M_D\,$,\,
 \,$\left<\phi\right>=174\,\textrm{GeV}$,\,
 \,$\left<\chi\right>=400\,\textrm{GeV}$\, and
 $\,(f,\,g,\,y)=(0.16,\,0.16,\,0.34)={O}(0.1)$,\,
 we can estimate the light neutrino mass scale,
 $\,\overline{m}_{II}^{}
 ={O}(m_{\nu}^{II})={O}(10\,m_{\nu}^{I})
 \simeq 0.05\,\textrm{eV}$.\,
 Subsequently, we derive,
 \,$B_{f}^{}B_{s}^{}
    \simeq 0.99\times 0.013\simeq 0.012$,\,
 and \,$K_{\eta}^{}\simeq 84$.\,
 This leads to,
 \,$\varepsilon_{}^{II} \simeq
    -2.3\times 10^{-5}_{}$\,,\,
 for the maximal CP phase.
 Using the approximate analytical
 formula\,\cite{kt1990,ht2001}
 for the baryon asymmetry, we arrive at
 \begin{eqnarray}
 \label{asymmetry} Y_{B}^{} ~\simeq~
 - \frac{28}{79} \times\frac{0.3
 \left(\varepsilon_{~}^{II}/g_{\ast}^{}\right)}
 {K_{\eta}^{}\left(\ln K_{\eta}^{}\right)^{0.6}_{}}
 ~\simeq~ 10^{-10}_{}\,,
 \end{eqnarray}
 consistent with the present observation\,\cite{pdg2006}.
 Furthermore, we note that in the Scheme-2,
 the active neutrino masses are dominated by
 the type-II Dirac seesaw,
 $\,m_{\nu}^{II}
 ={O}(0.1\, \textrm{eV})\gg m_{\nu}^{I}$,\,
 where both the NH and IH neutrino-mass-spectra
 can be realized.

 \vspace*{4mm}
 In this paper, we have presented a new possibility to
 realize the neutrinogenesis in the Dirac seesaw scenario.
 In our minimal construction, we introduce a real scalar
 $\chi$, a heavy singlet Dirac fermion $D$ and
 a heavy doublet scalar $\eta$ besides
 three right-handed singlet neutrinos to the SM.
 Therefore, different from previous
 realistic neutrinogenesis models,
 the \textit{radiative vertex corrections}
 rather than the self-energy corrections interfere
 with the tree-level diagrams to generate the CP
 asymmetry in the decays of the heavy particles.
 Finally, we note that the real singlet
 scalar $\chi$ at the weak scale
 can couple to the SM Higgs doublet $\phi$
 via the $Z_2^{}$-conserving quartic interaction
 \,$\chi\chi\phi^\dagger_{}\phi\,$.\,
 In consequence, the lightest neutral Higgs boson
 $h^0_{}$ is
 a mixture between \,$\phi^0_{}$\, and \,$\chi$\,,\,
 leading to non-SM-like anomalous couplings of $h^0_{}$
 with the weak gauge bosons ($W^\pm,\,Z$) and
 the SM-fermions. This can significantly modify
 the light Higgs boson ($h^0_{}$) phenomenology at
 the Tevatron Run-2, the CERN LHC and
 the future International Linear Colliders\,(ILC)
 \cite{bgm1977}.
 A systematical study for the collider phenomenology
 of $\phi^0_{}$ and $\chi$ is beyond the present scope
 and will be given elsewhere.

\end{document}